# Few-Shot Speaker Identification Using Depthwise Separable Convolutional Network with Channel Attention

*Yanxiong Li[1], Wucheng Wang[1], Hao Chen[1], Wenchang Cao[1], Wei Li[2], Qianhua He[1]*

[1]School of Electronic and Information Engineering, South China University of Technology, Guangzhou, China
[2]Computer Engineering Technical College, Guangdong Polytechnic of Science and Technology, Zhuhai, China

eeyxli@scut.edu.cn

## Abstract

Although few-shot learning has attracted much attention from the fields of image and audio classification, few efforts have been made on few-shot speaker identification. In the task of few-shot learning, overfitting is a tough problem mainly due to the mismatch between training and testing conditions. In this paper, we propose a few-shot speaker identification method which can alleviate the overfitting problem. In the proposed method, the model of a depthwise separable convolutional network with channel attention is trained with a prototypical loss function. Experimental datasets are extracted from three public speech corpora: Aishell-2, VoxCeleb1 and TORGO. Experimental results show that the proposed method exceeds state-of-the-art methods for few-shot speaker identification in terms of accuracy and F-score.

## 1. Introduction

In some practical applications, it is difficult to collect enough training speech samples which have the same distribution as testing speech samples for the task of speaker recognition. For example, the law enforcement agencies cannot always obtain adequate training speech samples uttered by the criminals for forensic speaker identification [1]. Building a reliable system for speaker identification with few utterances is still a challenging problem [2]-[5] and thus needs to be further investigated. This problem can be called *few-shot speaker identification* (FSSI), in which not-seen speakers (from the testing dataset) need to be identified by the model which has been trained by another speech corpus (the training dataset) [3]. What's more, the model is required to generalize well with a few speech samples for speakers not seen during the training stage. The task of FSSI requires to use several speech samples from testing domain in enrollment and testing stages, but the speakers of testing data are different from that of training data. In the task of FSSI, there are mismatches between training and testing datasets, including the mismatches of speakers' identities and numbers, speaking places, recording devices, etc. In addition, the model is easy to overfit the training datasets in few-shot learning [6].

Inspired by the success of few-shot learning for video and audio classification [7]-[9], we propose a FSSI method in this paper. To be specific, backbone network of the proposed framework is a depthwise separable convolutional network (DSCN) with channel attention (CA). On the one hand, the DSCN is a lightweight convolutional neural network (CNN) with very few parameters, and thus it has the potential to alleviate the overfitting problem [10]. On the other hand, in the DSCN, a standard convolution is factorized into a depthwise convolution and a $1\times1$ pointwise convolution without interaction among various channels [10]. Therefore, the DSCN is not good at utilizing channel information which can be effectively utilized by the CA module [11]. The study in this paper is the first work to adopt the DSCN with CA for FSSI. In conclusion, the contributions of this work are as follows:

1) We propose a FSSI method using a DSCN with CA for alleviating the overfitting problem.
2) We compare our method with previous methods on experimental datasets selected from three speech corpora and obtain state-of-the-art FSSI performance.

The rest of this paper is structured as follows. Section 2 briefly introduces related works, and Section 3 describes the proposed method in detail. Experiments are presented in Section 4, and conclusions are finally drawn in Section 5.

## 2. Related Works

Speaker recognition mainly includes two subtasks [12]: speaker identification [13]-[19] and speaker verification [20]-[31]. The former aims to determine which registered speaker provides a given utterance from a set of known speakers, while the latter aims to accept or reject the identity claim of a speaker. The work discussed in this paper focuses on the task of speaker identification.

The efforts in the works [13]-[48] mainly focused on two questions: to learn or design an effective front-end feature and to build a back-end model with stronger discriminative ability. Many kinds of hand-crafted features are designed to represent speaker's time-frequency properties, mainly including Mel-frequency cepstral coefficients [32], constant Q cepstral coefficients [33], linear prediction coding coefficients [32], eigenvoice-motivated vectors [34], and I-vector [35], [36]. Hand-crafted features are generally designed for specific situations. Therefore, they lack the ability of generalization. What's more, they are shallow instead of deep-model-based features. As a result, they cannot deeply characterize the time-frequency differences among various speakers. To overcome their deficiencies, many deep-model-based features have been proposed. These transformed features are proved to perform better than the hand-crafted features, mainly including the X-vector learned by a time-delay neural network [37], [38]; the speaker embeddings learned by: a CNN [39], a long short-term

Corresponding author: Yanxiong Li (eeyxli@scut.edu.cn).
This work was supported by national natural science foundation of China (62111530145, 61771200), international scientific research collaboration project of Guangdong Province (2021A0505030003), Guangdong basic and applied basic research foundation (2021A1515011454), and special project in key field of "Artificial Intelligence" in colleges and universities of Department of Education of Guangdong Province (2019KZDZX1045).

memory network [40] or a Siamese neural network (SNN) [41], [42]. On the other hand, the common back-end model mainly includes probabilistic linear discriminant analysis (PLDA) [20]-[30], Gaussian mixture model [43], hidden Markov model [44], vector quantization [45], support vector machine [46], and deep neural networks [47], [48].

Almost all of the works introduced above assume that speech samples are enough for training the models. When the training speech samples are few, the methods proposed in these works might not be so effective. To overcome the negative impact of few speech samples to the performance of speaker recognition methods, some researchers attempt to do works related to few-shot speaker recognition [2]-[5]. The works in [2]-[5] are directly related to the work in this paper and thus are further introduced in the following paragraphs.

Mishra [2] applies a SNN [49] to speaker verification and designs some loss functions, including center loss, categorical cross-entropy loss, and speaker-bias loss. The author uses a 3D-CNN for obtaining performance improvement on the task of text-independent speaker verification.

Wang et al. [3] proposes a speaker recognition method using centroid-based deep metric learning. They compare the prototypical loss with the triplet loss by organizing experiments for speaker verification and identification. Their experiments show that the prototypical loss outperforms the triplet loss, especially in the condition of small-scale training dataset.

Kye et al [4] propose a strategy for alleviating the negative influence of short utterances on the performance of speaker recognition by meta-learning. To improve the performance on various length of utterances, support set includes long utterances only, while query set consists of utterances with various durations. Besides, a prototypical network [50] is trained and enhanced by cross training.

In [5], an auto-encoder [51], [52] is combined with a capsule network [53] to learn generalized speaker embedding. In addition, the prototypical loss [50], instead of the common cross-entropy loss, is used to optimize speaker embedding. The method which is presented in [5], obtains an improvement for text-independent speaker identification in comparison with the baseline methods.

In summary, the works presented in [2]-[5] definitely contribute to the development of few-shot speaker recognition. However, almost all of them adopt the same speech corpus for selecting their training and testing datasets and standard convolution is adopted in the convolutional blocks. That is, there is almost no mismatch between the training dataset and the testing dataset in their experiments. Hence, the overfitting problem for FFSI caused by the mismatch between training and testing conditions is not explicitly considered in these works.

## 3. Method

In this section, we describe the proposed method in detail, mainly including the speaker embedding learning and loss function calculation.

### 3.1. Speaker Embedding Learning

As shown in Figure 1, the proposed framework for speaker embedding learning consists of the following parts: depthwise separable convolution (DSC), CA, and output modules. The motivation for the usage of the combination of DSC with CA modules is based on two considerations. First, the DSC can greatly reduce the size of the model (network) for making the model more concise. As a result, the model will have stronger generalization ability from the training condition to the testing condition. That is, the usage of the DSC module helps the model alleviate the overfitting problem. Second, the CA module can make the model focus on the important channel information which is not effectively utilized in the DSC module. Hence, the combination of the DSC module with the CA module is expected to be able to alleviate the overfitting problem with better performance for FSSI through learning a discriminative speaker embedding from each speech sample.

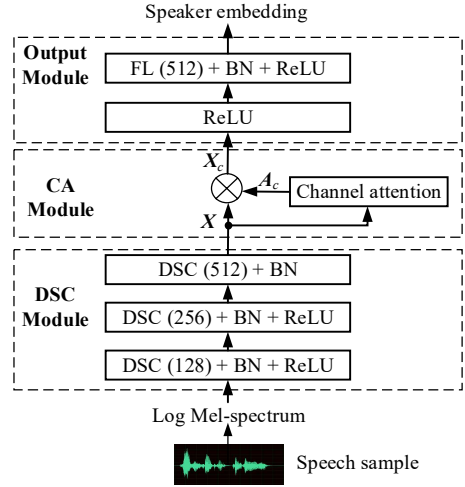

Figure 1: *The proposed framework for speaker embedding learning. BN, ReLU and FL denote batch normalization, rectified linear unit and fully connected layer, respectively.*

The framework depicted in Figure 1 works as follows. First, the audio feature of Log Mel-spectrum [54], [55] is extracted from each speech sample, and then is fed to the DSC module. Afterwards, the DSC and CA modules are adopted for deep feature transformation. Finally, speaker embedding is obtained from the fully connected layer of the output module.

### 3.1.1. DSC Module

The DSC module includes 3 DSC blocks as shown in Figure 1. Each DSC block consists of the following operations: DSC, batch normalization (BN), and/or rectified linear unit (ReLU). The digits (e.g., 512) in the parentheses denote number of channels. Specific structure of a DSC block is shown in Figure 2. The DSC block includes two steps: depthwise convolution and $1\times 1$ pointwise convolution [56]. Depthwise convolution uses different two-dimensional kernels to process each channel feature map, and is computationally efficient in comparison with standard convolution. However, depthwise convolution only filters input channels without combining them to generate new feature maps. Pointwise convolution is used to compute a linear combination of the output of depthwise convolution by $1\times 1$ convolution for producing new feature maps. Pointwise convolution uses $1\times 1\times m$ kernels to learn each channel feature and change the channel dimension. Compared to standard convolution kernel, DSC learns spatial feature maps and channel feature maps independently. With the same dimensions of input and output feature maps, DSC reduces Number of

Parameters (NoP) and computational cost of the network [56].

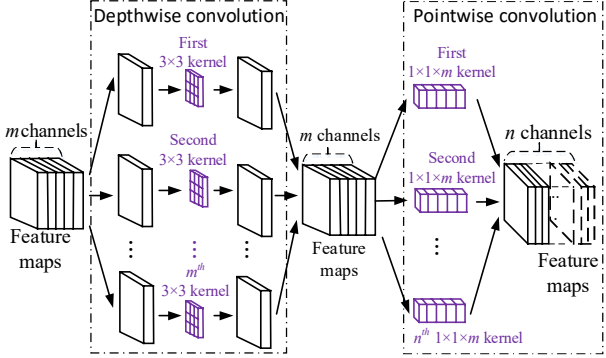

Figure 2: *The specific structure of a DSC block.*

Although the DSC can make the DSCN more efficient with respect to NoP and computational speed, the different channels considered in the DSCN do not interact with each other. As a result, channel information is not effectively utilized in the DSC module. To overcome this shortcoming, a CA module is added into the DSCN.

### 3.1.2. CA Module

As illustrated in Figure 3, the CA module consists of a maximum pooling operation, an average pooing operation, a feedforward network, and an element-wise summation & Sigmoid operation.

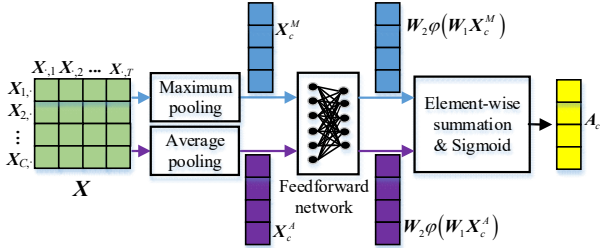

Figure 3: *Illustration of channel attention module.*

Let the input feature be $X \in \mathbf{R}^{C \times T}$, where $C$ and $T$ denote total number of channels and total number of frames, respectively. The temporal information in each channel is aggregated by average pooling and maximum pooling on time dimension. Let the outputs of average pooling and maximum pooling be $X_c^A = \{X_c^A\} \in \mathbf{R}^{C \times 1}$ and $X_c^M = \{X_c^M\} \in \mathbf{R}^{C \times 1}$, respectively, where $c$ is channel index and $1 \leq c \leq C$. $X_c^A$ and $X_c^M$ are defined by (1) and (2), respectively.

$$X_c^A = \frac{1}{T}\sum_{t=1}^{T} X_c(t), \quad (1)$$

$$X_c^M = \max(X_c(1), X_c(2), ..., X_c(T)). \quad (2)$$

Afterwards, the results of average pooling and maximum pooling are sequentially and independently fed to a feedforward network with two fully connected layers. The coefficient vector, $A_c$, is obtained via fusing the network's outputs by element-wise summation:

$$A_c = \sigma(W_2\varphi(W_1 X_c^A) + W_2\varphi(W_1 X_c^M)), \quad (3)$$

where $W_1$ and $W_2$ are weight matrices of the two layers of the feedforward network; $\sigma(\cdot)$ and $\varphi(\cdot)$ are the activation functions of Sigmoid and ReLU, respectively. Finally, the output of channel attention module, $X_c$, is defined by the element-wise multiplication of the feature $X$ and the coefficient vector $A_c$:

$$X_c = A_c \otimes X, \quad (4)$$

where $\otimes$ represents the element-wise multiplication.

In summary, CA module is used to obtain coefficient vector for channel attention, which makes up for the disadvantage of the DSC (not effectively utilize channel information). In addition, the usage of channel attention almost does not increase the NoP of the network. Hence, the proposed DSCN with CA is expected to be able to alleviate the overfitting problem with better performance.

### 3.1.3. Output Module

The output module consists of a ReLU layer and a fully connected layer with operations of both BN and ReLU. From Figure 3 and Eq. (1) to Eq. (4), it can be known that the CA module outputs a tensor (i.e., $X_c$) of dimension $C \times T$. As shown in Figure 1, the output of the CA module is fed to the output module where the layer of "FL (512) + BN + ReLU" converts the tensor of dimension $C \times T$ into a 1-dimensional tensor (i.e., the learned speaker embedding).

## 3.2. Loss Function Calculation

Prototypical loss function is adopted to train the proposed framework since only a few training samples are required for generating class centers in training stage and a speaker identity can be produced based on few utterances in enrollment stage. It is defined by

$$Loss = -\log \frac{\exp(-dist(f_c(X), X_k))}{\sum_{k'} \exp(-dist(f_c(X), X_{k'}))} \quad (5)$$

where $dist(\cdot)$ is a distance function (e.g., Euclidean distance); $f_c(X)$ is a center vector; $X_k$ and $X_{k'}$ are speaker embeddings in query set. The calculation of loss function is illustrated in Figure 4, where the module of "speaker embedding learning" is implemented by the framework depicted in Figure 1.

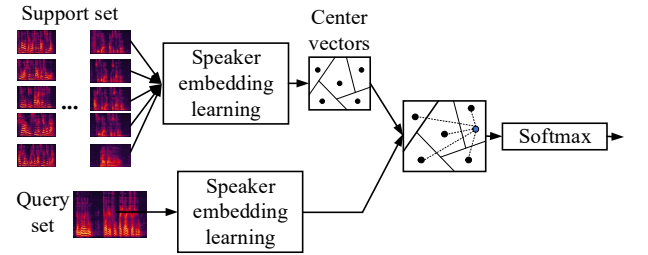

Figure 4: *Illustration of the calculation of loss function.*

We first use training dataset to generate support set and query set and then to conduct *K*-way *N*-shot learning (*K*=5, *N*=5). During the training stage, speech samples of both support set and query set are fed to the proposed framework for learning speaker embedding. In the support set, all speaker embeddings from the same speaker are averaged as a center vector. In the query set, each speaker embedding is compared to all center vectors by a distance function for calculating the prototypical loss function. During the enrollment stage, few utterances are adopted to generate center vectors. During the testing stage, the distance between the speaker embedding of one testing sample and one of all center vectors are sequentially calculated. The enrolled speaker corresponding to the center vector with the minimum distance is the speaker identification result of the testing speech sample.

## 4. Experiments

This section describes experimental datasets, setup, and results.

### 4.1. Experimental Datasets

Experiments are conducted on the datasets selected from three public speech corpora: Aishell-2 [57], VoxCeleb1 [58], and TORGO [59]. Aishell-2 includes 1000-hour Mandarin speech recordings uttered by 1991 speakers. Its speech recordings are acquired in quiet environment with IOS-system mobile phone (16 kHz, 16 bit). VoxCeleb1 includes more than 100 thousand speech recordings uttered by 1251 speakers, with total length of 352 hours. Its speech recordings are acquired in noisy environment. TORGO is a pathological speech corpus uttered by 7 normal speakers and 8 dysarthria patients. It is difficult to collect utterances from patients. Therefore, TORGO is a small-scale corpus compared to other two speech corpora.

Both Aishell-2 and VoxCeleb1 are independently divided into three datasets: training, enrollment, and testing datasets. Since TORGO is a small-scale corpus compared to Aishell-2 and VoxCeleb1, it is used only for generating enrollment and testing datasets. In the training dataset, 80 speakers are randomly selected from Aishell-2 and VoxCeleb1, and each speaker has 200 speech samples. In the enrollment and testing datasets, 80, 80 and 15 speakers are selected from Aishell-2, VoxCeleb1 and TORGO, respectively; and each speaker has 20 and 50 speech samples in the enrollment and testing datasets, respectively. The length of speech samples in the experimental datasets is from 2 s to 5 s. When we select experimental data from the speech corpora in each time, the speakers and speech samples per speaker in the training, enrollment and testing datasets are different to each other. When the training dataset is selected from Aishell-2, the enrollment and testing datasets are chosen from VoxCeleb1. When the enrollment and testing datasets are chosen from Aishell-2 and TORGO, the training dataset is selected from VoxCeleb1. That is, the training and testing datasets are from different speech corpora and thus there is mismatch between the training dataset and the testing dataset.

### 4.2. Experimental Setup

The Kaldi and PyTorch toolkits are used to extract audio features and to train network, respectively. Experiments are conducted on a machine whose main configurations are: four NVIDIA 1080TI GPU, one Intel CPU i7-6700 with 3.10 GHz, and 48 GB RAM.

The Log Mel-spectrum with 39-dimension is extracted from each speech sample. Frame length and shift are 25 ms and 12.5 ms, respectively. Three DSC blocks with numbers of channels of 128, 256, and 512 are sequentially stacked. The operations of ReLU function and BN are successively conducted on the output of each DSC layer. The maximum epoch for network training is 80, and the learning rate is 0.0001. The performance metrics are accuracy and F-score, which are introduced in detail in [60]. The higher the accuracy and F-score are, the better the performance of the method is.

Training, enrollment and testing datasets are independently adopted to generate support and query sets for conducting the 5-way 5-shot learning. For example, in the training dataset, we randomly choose 5 speakers (5-way) from 80 speakers and select 5 speech samples (5-shot) from 200 speech samples per speaker to generate support set or query set. The selection procedures of speakers and speech samples per speaker are repeated many times (e.g., 50 times) for covering all speakers and their speech samples in the training dataset (or enrollment, testing datasets). The selected speakers and speech samples in different times are different to each other. The average score of the repeated results is used as the final result.

### 4.3. Experimental Results

We first discuss the performance of the framework with DSC, SC (standard convolution) and CA modules, and then present the performance comparison of different methods.

*4.3.1. Ablation Experiments*

To prove the contributions of the SC, DSC and CA modules to the performance of the framework (depicted in Figure 1), four ablation experiments are conducted independently, including the frameworks with: SC module, DSC module, both SC and CA modules, and both DSC and CA modules. The size of the framework with the same modules (e.g., SC, DSC) is the same when the framework is evaluated on experimental datasets. The metric of NoP is used to measure the framework's spatial complexity. The results of ablation experiments are presented in Table 1.

Table 1: *Results of ablation experiments*.

| Frameworks with | Metrics | Aishell-2 | VoxCeleb1 | TORGO |
|---|---|---|---|---|
| SC | NoP | | 1985303 | |
| | Accuracy (%) | 91.34 | 81.66 | 84.37 |
| | F-score (%) | 92.97 | 83.52 | 85.29 |
| DSC | NoP | | 1443728 | |
| | Accuracy (%) | 86.71 | 77.21 | 80.68 |
| | F-score (%) | 88.35 | 79.65 | 82.71 |
| SC and CA | NoP | | 2013716 | |
| | Accuracy (%) | 94.07 | 85.76 | 88.74 |
| | F-score (%) | 95.73 | 88.35 | 90.08 |
| DSC and CA | NoP | | 1464458 | |
| | Accuracy (%) | 94.46 | 86.42 | 89.24 |
| | F-score (%) | 96.18 | 88.74 | 90.62 |

NoP: Number of Parameters.

Based on the results in Table 1, the following three observations can be obtained.

First, the framework with DSC module does not exceed the framework with SC module in terms of both accuracy and F-score. The reason is that the channel information is not effectively utilized during the procedure of convolutional calculation in the DSC module, whereas the channel information is used effectively in the SC module. On the other hand, the NoP (i.e., 1443728) of the framework with DSC module is much less than that (i.e., 1985303) of the framework with SC module. In summary, compared to the SC module, the shortcoming of the DSC module is that it cannot effectively utilize the channel information, which results in performance degradation regarding accuracy and F-score. Its advantage over the SC module is that the size of the framework is greatly reduced due to the usage of DSC.

Second, the framework with both DSC and CA modules outperforms the framework with both SC and CA modules in terms of accuracy, F-score, and NoP. The reasons are that the channel information is effectively utilized by the CA module and the overfitting problem is alleviated by the DSC module. Therefore, the framework with both DSC and CA modules exceeds other frameworks, including the framework with both SC and CA modules. The framework with both DSC and CA modules is the proposed neural network architecture with NoP of 1464458, whereas the framework with both SC and CA modules is the baseline neural network architecture with NoP of 2013716. That is, the proposed neural network architecture has less parameters, and can be regarded as a compressed version of the baseline neural network architecture in terms of NoP. What's more, the proposed neural network architecture has advantage over the baseline neural network architecture in terms of accuracy and F-score. In summary, the proposed neural network architecture has less parameters and has better generalization ability compared to the baseline neural network architecture.

Third, the framework with both SC and CA modules obtains higher scores of both accuracy and F-score than the framework with SC module, and the framework with both DSC and CA modules also achieves higher scores of both accuracy and F-score than the framework with DSC module. That is, the CA module is beneficial for improving the performance of the framework in terms of both accuracy and F-score when evaluated on all experimental datasets. For example, the framework with both SC and CA modules produces accuracy of 94.07% and F-score of 95.73% on the datasets from Aishell-2, and obtains accuracy improvement by 2.73% (94.07% - 91.34%) and F-score improvement by 2.76% (95.73% - 92.97%), compared with the framework with SC module. The possible reasons are that the information of critical channels is effectively utilized by the CA module. On the other hand, adding the CA module into the framework increases the size of the framework. However, the increment of the NoP is relatively small. For example, the NoP of the framework with SC module is 1985303, while the NoP of the framework with SC and CA modules is 2013716.

*4.3.2. Comparison of Different Methods*

In addition to the ablation experiments above, we carry out comparative experiments between the proposed method and state-of-the-art methods. Main parameters of previous methods are set according to the suggestions in corresponding references and are optimally tuned on the experimental datasets adopted in this work. These previous methods can be divided into three categories: shallow-model based method, deep-model based method, and few-shot-learning (metric-learning) based method. The typical shallow-model based method is the I-vector + PLDA [61]. The deep-model based method includes the X-vector + PLDA [62] and the D-vector + PLDA [63]. Few-shot-learning based methods include the methods based on prototypical network [50], Siamese network [49], matching network [64], and relation network [65]. Under the same experimental conditions, the results obtained by different methods are listed in Table 2.

Table 2: *Results obtained by different methods (in %).*

| Methods | Aishell-2 | | VoxCeleb1 | | TORGO | |
|---|---|---|---|---|---|---|
| | *Acc.* | *F*-score | *Acc.* | *F*-score | *Acc.* | *F*-score |
| I + P [61] | 79.60 | 78.51 | 65.16 | 64.58 | 65.79 | 64.71 |
| X + P [62] | 71.31 | 68.56 | 61.35 | 58.38 | 62.79 | 60.65 |
| D + P [63] | 55.71 | 52.38 | 39.56 | 36.32 | 43.48 | 41.63 |
| PN [50] | 91.34 | 92.97 | 82.66 | 83.52 | 83.52 | 85.29 |
| SN [49] | 72.95 | 73.17 | 68.74 | 70.13 | 70.24 | 71.25 |
| MN [64] | 81.42 | 78.31 | 68.61 | 66.40 | 70.54 | 68.85 |
| RN [65] | 89.46 | 88.10 | 79.39 | 78.04 | 82.81 | 80.53 |
| **Ours** | **94.46** | **96.18** | **86.42** | **88.74** | **89.24** | **90.62** |

I + P: I-vector + PLDA; X + P: X-vector + PLDA; D + P: D-vector + PLDA; PN: prototypical network; SN: Siamese network; MN: matching network; RN: relation network; Acc.: Accuracy.

It can be seen from Table 2 that our method outperforms all previous methods on the experimental datasets selected from three speech corpora in terms of both accuracy and F-score. The reasons are explained as follows.

First, the models in both the shallow-model based method and the deep-model based methods tend to remember the identity information of specific speakers in training speech samples, rather than the information of characteristic difference among all speakers. As a result, in the task of FSSI, both the shallow-model and deep-model based methods cannot be effectively generalized very well from the training speech samples to the testing speech samples.

Second, although the state-of-the-art FSSI methods have potential capacities to alleviate the problem of mismatch between training and testing conditions, they did not take explicit measures for tackling the overfitting problem, such as the usage of both DSC and CA modules in our work. Hence, the previous few-shot-learning based methods do not achieve the most satisfactory results for FSSI, although they almost always outperform both the shallow-model based method and the deep-model based methods.

Finally, all methods achieve better results when they are evaluated on the experimental datasets selected from the Aishell-2, and obtain worse results when they are assessed on the experimental datasets selected from the VoxCeleb1. The reason is probably that speech samples selected from the Aishell-2 are recorded in quiet environment without background noises, while speech samples selected from the VoxCeleb1 are degraded by relatively strong real-world noises, including background utterances (such as chatter and laughter), room reverberations, noises of recording device.

# 5. Conclusions

In this paper, we propose a method of FSSI using a DSCN with CA. Experimental datasets are selected from three public speech corpora, and the results show that the proposed method outperforms state-of-the-art methods for FSSI. Based on the experimental results, we conclude that the combination of DSC with CA modules in the proposed framework is effective for performance improvement and is also beneficial for alleviating the overfitting problem for FSSI.

This study is a preliminary work, and there is still room for improvement in terms of accuracy, F-score and NoP. In the next work, we plan to further optimize our method via integrating other techniques into the proposed framework, such as the usage of capsule network, self-attention mechanism, other types of loss functions.

# 6. References

[1] J.P. Campbell, W. Shen, W.M. Campbell, R. Schwartz, J.F. Bonastre, and D. Matrouf, "Forensic speaker recognition," *IEEE SPM*, vol. 26, no. 2, pp. 95-103, 2009.

[2] P. Mishra, "Few shot text-independent speaker verification using 3D-CNN," *arXiv preprint arXiv:2008.11088*, 2020, pp. 1-7.

[3] J. Wang, K. Wang, M. T. Law, F. Rudzicz, and M. Brudno, "Centroid-based deep metric learning for speaker recognition," in *Proc. of IEEE ICASSP*, 2019, pp. 3652-3656.

[4] S.M. Kye, Y. Jung, H.B. Lee, S.J. Hwang, and H. Kim, "Meta-learning for short utterance speaker recognition with imbalance length pairs," *arXiv preprint arXiv:2004.02863*, 2020, pp. 1-5.

[5] P. Anand, A.K. Singh, S. Srivastava, and B. Lall, "Few shot speaker recognition using deep neural networks," *arXiv preprint arXiv:1904.08775*, 2019, pp. 1-5.

[6] O. Klejch, J. Fainberg, and P. Bell, "Learning to adapt: a meta-learning approach for speaker adaptation," *arXiv preprint arXiv:1808.10239*, 2018, pp. 1-5.

[7] L. Zhu, and Y. Yang, "Label independent memory for semi-supervised few-shot video classification," *IEEE TPAMI*, vol. 44, no. 1, pp. 273-285, 2022.

[8] B. Shi, M. Sun, K.C. Puvvada, C.C. Kao, S. Matsoukas, and C. Wang, "Few-shot acoustic event detection via meta learning," in *Proc. of IEEE ICASSP*, 2020, pp. 76-80.

[9] Y. Wang, N.J. Bryan, M. Cartwright, J.P. Bello, and J. Salamon, "Few-shot continual learning for audio classification," in *Proc. of IEEE ICASSP*, 2021, pp. 321-325.

[10] F. Chollet, "Xception: deep learning with depthwise separable convolutions," in *Proc. of IEEE CVPR*, 2017, pp. 1800-1807.

[11] M. Choi, H. Kim, B. Han, N. Xu, and K.M. Lee, "Channel attention is all you need for video frame interpolation," in *Proc. of the AAAI Conference on Artificial Intelligence*, 2020, pp. 10663-10671.

[12] C.S. Greenberg, L.P. Mason, S.O. Sadjadi, and D.A. Reynolds, "Two decades of speaker recognition evaluation at the national institute of standards and technology," *Comput. Speech Lang.*, vol. 60, pp. 1-10, 2020.

[13] F. Alegre, G. Soldi, N. Evans, B. Fauve, and J. Liu, "Evasion and obfuscation in speaker recognition surveillance and forensics," in *Proc. of IWBF*, 2014, pp. 1-6.

[14] M.T.S. Al-Kaltakchi, W.L. Woo, S.S. Dlay, and J.A. Chambers, "Study of fusion strategies and exploiting the combination of MFCC and PNCC features for robust biometric speaker identification," in *Proc. of ICBF*, 2016, pp. 1-6.

[15] M.E. Ayadi, A.K.S.O. Hassan, A. Abdel-Naby, and O.A. Elgendy, "Text-independent speaker identification using robust statistics estimation," *Speech Commun.*, vol. 92, pp. 52-63, 2017.

[16] X. Wang, F. Xue, W. Wang, and A. Liu, "A network model of speaker identification with new feature extraction methods and asymmetric BLSTM," *Neurocomputing*, vol. 403, pp. 167-181, 2020.

[17] A. Nelus, S. Rech, T. Koppelmann, H. Biermann, and R. Martin, "Privacy-preserving Siamese feature extraction for gender recognition versus speaker identification," in *Proc. of INTERSPEECH*, 2019, pp. 3705-3709.

[18] J. Malek, J. Jansky, T. Kounovsky, Z. Koldovsky, and J. Zdansky, "Blind extraction of moving audio source in a challenging environment supported by speaker identification via X-vectors," in *Proc. of IEEE ICASSP*, 2021, pp. 226-230.

[19] Z. Li, and J. Whitehill, "Compositional embedding models for speaker identification and diarization with simultaneous speech from 2+ speakers," in *Proc. of IEEE ICASSP*, 2021, pp. 7163-7167.

[20] A. Roy, M. Magimai-Doss, and S. Marcel, "A fast parts-based approach to speaker verification using boosted slice classifiers," *IEEE TIFS*, vol. 7, no. 1, pp. 241-254, 2012.

[21] A. Larcher, K.A. Lee, B. Ma, and H. Li, "Text-dependent speaker verification: Classifiers, databases and RSR2015," *Speech Commun.*, vol. 60, pp. 56-787, 2014.

[22] B.C. Haris, and R. Sinha, "Robust speaker verification with joint sparse coding over learned dictionaries," *IEEE TIFS*, vol. 10, no. 10, pp. 2143-2157, 2015.

[23] J. Guo, N. Xu, K. Qian, Y. Shi, K. Xu, Y. Wu, and A. Alwan, "Deep neural network based I-vector mapping for speaker verification using short utterances," *Speech Commun.*, vol. 105, pp. 92-102, 2018.

[24] A. Poddar, M. Sahidullah, and G. Saha, "Quality measures for speaker verification with short utterances," *Digit. Signal Process.*, vol. 88, pp. 66-79, 2019.

[25] Y. Wu, C. Guo, H. Gao, J. Xu, and G. Bai, "Dilated residual networks with multi-level attention for speaker verification," *Neurocomputing*, vol. 412, pp. 177-186, 2020.

[26] Z. Wang, W. Xia, and J.H.L. Hansen, "Cross-domain adaptation with discrepancy minimization for text-independent forensic speaker verification," in *Proc. of INTERSPEECH*, 2020, pp. 2257-2261.

[27] M. Sang, W. Xia, and J.H.L. Hansen, "Open-set short utterance forensic speaker verification using teacher-student network with explicit inductive bias," in *Proc. of INTERSPEECH*, 2020, pp. 2262-2266.

[28] L. Zheng, J. Li, M. Sun, X. Zhang, and T.F. Zheng, "When automatic voice disguise meets automatic speaker verification," *IEEE TIFS*, vol. 16, pp. 824-837, 2021.

[29] A. Gomez-Alanis, J.A. Gonzalez-Lopez, S.P. Dubagunta, A.M. Peinado, and M. Magimai-Doss, "On joint optimization of automatic speaker verification and anti-spoofing in the embedding space," *IEEE TIFS*, vol. 16, pp. 1579-1593, 2021.

[30] Y. Zhang, M. Yu, N. Li, C. Yu, J. Cui, and D. Yu, "Seq2Seq attentional Siamese neural networks for text-dependent speaker verification," in *Proc. of IEEE ICASSP*, 2019, pp. 6131-6135.

[31] A. Hajavi, and A. Etemad "Siamese capsule network for end-to-end speaker recognition in the wild," in *Proc. of IEEE ICASSP* 2021, pp. 7203-7207.

[32] A. Chowdhury, and A. Ross, "Fusing MFCC and LPC features using 1D triplet CNN for speaker recognition in severely degraded audio signals," *IEEE TIFS*, vol. 15, pp. 1616-1629, 2020.

[33] M. Todisco, H. Delgado, and N. Evans, "A new feature for automatic speaker verification anti-spoofing: constant Q cepstral coefficients," in *Proc. of Odyssey*, 2016, pp. 283-290.

[34] P. Kenny, G. Boulianne, and P. Dumouchel, "Eigenvoice modeling with sparse training data," *IEEE TSAP*, vol. 13, no. 3, pp. 345-354, 2005.

[35] N. Dehak, P. J. Kenny, et al, "Front-end factor analysis for speaker verification," *IEEE TASLP*, vol.19, no. 4, pp. 788-798, 2011.

[36] U. Khan, M. India, and J. Hernando, "I-vector transformation using k-nearest neighbors for speaker verification," in *Proc. of IEEE ICASSP*, 2020, pp. 7574-7578.

[37] D. Snyder, D. Garcia-Romero, G. Sell, D. Povey, and S. Khudanpur, "X-vectors: Robust DNN embeddings for speaker recognition," in *Proc. of IEEE ICASSP*, 2018, pp. 5329-5333.

[38] D. Snyder, D. Garcia-Romero, G. Sell, A. McCree, D. Povey, and S. Khudanpur, "Speaker recognition for multi-speaker conversations using X-vectors," in *Proc. of IEEE ICASSP*, 2019, pp. 5796-5800.

[39] Y. Li, W. Wang, M. Liu, Z. Jiang, and Q. He, "Speaker clustering by co-optimizing deep representation learning and cluster estimation," *IEEE TMM*, vol. 23, pp. 3377-3387, 2021.

[40] L. Wan, Q. Wang, A. Papir, and I.L. Moreno, "Generalized end-to-end loss for speaker verification," in *Proc. of IEEE ICASSP*, 2018, pp. 4879-4883.

[41] S. Soleymani, A. Dabouei, S.M. Iranmanesh, H. Kazemi, J. Dawson, and N.M. Nasrabadi, "Prosodic-enhanced Siamese convolutional neural networks for cross-device text-independent speaker verification," in *Proc. of IEEE BTAS*, 2018, pp. 1-7.

[42] U. Khan, and J. Hernando, "Unsupervised training of Siamese networks for speaker verification," in *Proc. of INTERSPEECH*, 2020, pp. 3002-3006.

[43] B. Narayanaswamy, and R. Gangadharaiah, "Extracting additional information from Gaussian mixture model probabilities for

improved text independent speaker identification," in *Proc. of IEEE ICASSP*, 2005, pp. I/621-I/624.
[44] S. Ismail, "Enhancing speaker identification performance under the shouted talking condition using second-order circular hidden Markov models," *Speech Commun.*, vol. 48, no. 8, pp. 1047-1055, 2006.
[45] P.C. Nguyen, M. Akagi, and T.B. Ho, "Temporal decomposition: a promising approach to VQ-based speaker identification," in *Proc. of IEEE ICASSP*, 2003, pp. III-617-620.
[46] I. Zeljkovic, P. Haffner, B. Amento, and J. Wilpon, "GMM/SVM N-best speaker identification under mismatch channel conditions," in *Proc. of IEEE ICASSP*, 2008, pp. 4129-4132.
[47] R. Jahangir, et al., "Text-independent speaker identification through feature fusion and deep neural network," *IEEE Access*, vol. 8, pp. 32187-32202, 2020.
[48] Z. Bai, and X. Zhang, "Speaker recognition based on deep learning: An overview," *Neural Networks*, vol. 140, pp. 65-99, 2021.
[49] G. Koch, R. Zemel, and R. Salakhutdinov, "Siamese neural networks for one-shot image recognition," in *Proc. of ICML*, 2015. pp. 1-8.
[50] J. Snell, K. Swersky, and R.S. Zemel, "Prototypical networks for few-shot learning," *arXiv preprint arXiv:1703.05175,* 2017, pp. 1-13.
[51] Y. Wang, H. Yao, and S. Zhao, "Auto-encoder based dimensionality reduction," *Neurocomputing,* vol. 184, no. 5, pp. 232-242, 2016.
[52] S. Lange, and M. Riedmiller, "Deep auto-encoder neural networks in reinforcement learning," in *Proc. of IJCNN*, 2010, pp. 1-8.
[53] S. Sabour, N. Frosst, and G.E. Hinton, "Dynamic routing between capsules," *arXiv preprint* arXiv:*1710.09829,* 2017, pp. 1-11.
[54] Y. Li, M. Liu, W. Wang, Y. Zhang, and Q. He, "Acoustic scene clustering using joint optimization of deep embedding learning and clustering iteration," *IEEE TMM*, vol. 22, no. 6, pp. 1385-1394, 2020.
[55] Y. Li, M. Liu, K. Drossos, and T. Virtanen, "Sound event detection via dilated convolutional recurrent neural networks," *Proc. of IEEE ICASSP*, 2020, pp. 286-290.
[56] A.G. Howard, M. Zhu, B. Chen, D. Kalenichenko, W. Wang, T. Weyand, M. Andreetto, and H. Adam, "Mobilenets: Efficient convolutional neural networks for mobile vision applications," *arXiv preprint arXiv: 1704.04861*, 2017, pp. 1-9.
[57] J. Du, X. Na, X. Liu, and H. Bu, "Aishell-2: Transforming mandarin ASR research into industrial scale," *arXiv preprint arXiv:1808.10583,* 2018, pp. 1-3.
[58] A. Nagrani, J.S. Chung, W. Xie, and A. Zissermana, "Voxceleb: Large-scale speaker verification in the wild," *Comput. Speech Lang.*, vol. 60, pp. 1-15, 2020.
[59] N.J. Neethu, and S. Umesh, "Improving acoustic models in TORGO dysarthric speech database," *IEEE TNSRE*, vol. 26, no. 3, pp. 637-645, 2018.
[60] M. Sokolova, N. Japkowicz, and S. Szpakowicz, "Beyond accuracy, F-score and ROC: a family of discriminant measures for performance evaluation," in *Proc. of Australasian joint conference on artificial intelligence*, 2006, pp. 1015-1021.
[61] K.A. Lee, A. Larcher, C.H. You, B. Ma, and H. Li, "Multi-session PLDA scoring of I-vector for partially open-set speaker detection," in *Proc. of INTERSPEECH*, 2013, pp. 3651-3655.
[62] D. Snyder, D. Garcia-Romero, D. Povey, and S. Khudanpur, "Deep neural network embeddings for text-independent speaker verification," in *Proc. of INTERSPEECH*, 2017, pp. 999-1003.
[63] E. Variani, X. Lei, E. McDermott, I.L. Moreno, and J. Gonzalez-Dominguez, "Deep neural networks for small footprint text-dependent speaker verification," in *Proc. of IEEE ICASSP*, 2014, pp. 4052-4056.
[64] O. Vinyals, C. Blundell, T. Lillicrap, K. Kavukcuoglu, and D. Wierstra, "Matching networks for one shot learning," *arXiv preprint arXiv: 1606.04080*, 2016, pp. 1-12.
[65] F. Sung, Y. Yang, L. Zhang, T. Xiang, P. H. S. Torr, and T. M. Hospedales, "Learning to compare: relation network for few-shot learning," in *Proc. of IEEE/CVF CVPR*, 2018, pp. 1199-1208.